\documentclass[prx,twocolumn,aps,10pt]{revtex4-2}

\usepackage{graphicx}
\usepackage[caption=false]{subfig}
\usepackage{dcolumn}
\usepackage{bm}
\usepackage{amsmath,amssymb,amsfonts,xfrac,mathrsfs}
\usepackage{comment}
\usepackage{braket}
\usepackage{xcolor}
\usepackage{import}
\usepackage{hyperref}
\usepackage{orcidlink}

\usepackage{tikz}
\usetikzlibrary{calc, positioning, arrows.meta, fit, shapes}

\begin{document}

\title[Quantum solver for single-impurity Anderson models with particle-hole symmetry]{Quantum solver for single-impurity Anderson models\\with particle-hole symmetry}
\thanks{\textbf{Notice:} This manuscript has been authored by UT-Battelle, LLC, under contract DE-AC05-00OR22725 with the US Department of Energy (DOE). The US government retains, and the publisher, by accepting the article for publication, acknowledges that the US government retains a nonexclusive, paid-up, irrevocable, worldwide license to publish or reproduce the published form of this manuscript, or allow others to do so, for US government purposes. DOE will provide public access to these results of federally sponsored research in accordance with the DOE Public Access Plan (\href{http://energy.gov/downloads/doe-public-access-plan}{http://energy.gov/downloads/doe-public-access-plan}).}

\author{M. Karabin$^{1,2}$}
\email{mariia.karabin@mtsu.edu}
\author{T. Sohail$^3$}
\author{D. Bykov$^3$}
\author{E. A. Coello P\'erez$^3$\orcidlink{0000-0001-6074-0711}}
\email{coellopereea@ornl.gov}
\author{S. Ghosh$^3$\orcidlink{0000-0003-3800-5264}}
\author{M.~Gopalakrishnan~Meena$^3$\orcidlink{0000-0003-4048-4639}}
\author{S. Kim$^3$\orcidlink{0000-0001-5906-3004}}
\author{A. Shehata$^3$}
\author{I.-S. Suh$^3$\orcidlink{0000-0002-6923-6455}}
\author{H. Terletska$^{1,2}$}
\author{M. Eisenbach$^3$\orcidlink{0000-0001-8805-8327}}
\affiliation{$^1$Department of Physics and Astronomy, Middle Tennessee State University, Murfreesboro, TN, USA}
\affiliation{$^2$Quantum Research Interdisciplinary Science and Education (QRISE) Center, Middle Tennessee State University, Murfreesboro, TN, USA}
\affiliation{$^3$National Center for Computational Sciences, Oak Ridge National Laboratory, Oak Ridge, TN, USA}

\date{\today}

\begin{abstract}
Quantum embedding methods, such as dynamical mean-field theory (DMFT), provide a powerful framework for investigating strongly correlated materials. A central computational bottleneck in DMFT is in solving the Anderson impurity model (AIM), whose exact solution is classically intractable for large bath sizes. In this work, we benchmark a quantum-classical hybrid solver tailored for particle-hole symmetric AIMs, using the variational quantum eigensolver (VQE) to prepare the ground state of the model with shallow quantum circuits. The solver uses shallow quantum ans\"atze and one set of variational parameters to prepare the ground state and its particle and hole excitations, enabling the construction of the impurity Green's function through a continued-fraction expansion. We evaluate the performance of this approach across a few bath sizes and interaction strengths under noisy, shot-limited conditions. We compare three optimization routines (COBYLA, Adam, and L-BFGS-B) in terms of convergence and fidelity, assess the benefits of estimating a quantum-computed moment (QCM) correction to the variational energies, and benchmark the approach by comparing the density of states (DOS) computed from the impurity Green's function against that obtained using a classical pipeline. Our results demonstrate the feasibility of Green's function construction on near-term devices and establish practical benchmarks for quantum impurity solvers embedded within self-consistent DMFT loops.
\end{abstract}

\maketitle

\section{Introduction}
Chemistry and condensed matter physics are governed by the collective behavior of interacting electrons~\cite{fulde_1995}. Since the direct numerical solution of the Schr\"odinger equation is computationally intractable for all but the smallest systems, many different approaches have been devised that employ various approximations to reduce the computational complexity of solving the many-electron problem~\cite{Rossmannek_2023}. Among these, density functional theory (DFT)~\cite{hohenberg_1964, kohn_1965} is a widely used and popular method that, while theoretically exact, relies on approximations to the exchange-correlation functional. These approximations often fail to capture strong local correlations, especially in systems exhibiting Mott physics or dynamical electron localization. One method to reintroduce electron correlation contributions missed by these approximations augments this approach with dynamical mean-field theory (DMFT)~\cite{kotliar_g_2006, georges_a_1996, zgrid_2011,bauer_2016, rungger_2020}, which maps the lattice model onto a local Anderson impurity model (AIM)~\cite{anderson_pw_1961, keen_t_2020, ma_2020}, providing the mean-field that describes the electrons in a lattice. The impurity model captures local dynamical fluctuations and becomes exact in the limit of infinite spatial dimensions. This mapping has been highly instrumental in describing correlated electron materials ranging from transition metal oxides to heavy fermion systems.

Solving the AIM within DMFT typically requires computationally expensive techniques such as exact diagonalization, continuous-time quantum Monte Carlo, or numerical renormalization group. Although accurate, these classical solvers scale poorly with bath size and may struggle with real-frequency or sign problems~\cite{gull_e_2011, georges_a_1996}, limiting their application to more complex systems. Furthermore, these methods can also become computationally expensive at low temperatures~\cite{schollwock_u_2011, bulla_r_2008}, where quantum entanglement and non-perturbative effects like the Kondo phenomenon play a crucial role.

Quantum computing introduces novel approaches to solving impurity models with the potential to overcome some of these limitations, owing to its ability to efficiently represent large Hilbert spaces and capture strong electron correlations~\cite{mcardle_s_2020, bauer_b_2016}. Some of these approaches construct a continued fraction representation of the model's Green's function~\cite{greenediniz_g_2024, jamet_f_2021} from quantum states that represent the ground state of the AIM and the particle and hole excitations of such a state at the impurity sites. The limitation of circuit depth in current quantum hardware makes the optimization of a shallow parametrized quantum circuit (PQC) via the variational quantum eigensolver (VQE)~\cite{peruzzo_a_2014, McClean_2016} a common strategy to prepare the AIM ground state in the noisy intermediate-scale quantum era~\cite{Jaderberg_2020}.

In this work, we evaluate the performance of a quantum solver for AIMs with particle-hole symmetry that incorporates multiple quantum subroutines to construct the impurity Green's function. The focus on the half-filled case in the present work serves as an important benchmark for our approach as this case has been extensively studied in the context of understanding Mott insulators before its generalization for inclusion in a DFT+DMFT workflow. To solve these AIMs we introduce a minimal yet expressive set of variational ans\"atze capable of preparing the model's ground state and the particle and hole excitations needed to compute the impurity Green's function using a common set of variational parameters. This approach eliminates the computational overhead associated with the separate optimization of circuits preparing excited states~\cite{jamet_f_2021}, or the additional measurement costs incurred by methods that rely on mid-circuit measurement~\cite{jones_eb_2025}. We benchmark variational results against exact diagonalization to assess the accuracy and robustness of this ground-state preparation method. The corresponding excited states are generated via structured parameter shifts applied to the optimized ground-state circuit, avoiding the need to re-optimize each excited state. The performance of the proposed solver is evaluated by computing the impurity Green's functions for multiple values of the Hubbard interaction and bath sizes. Finally, we analyze the computational overhead associated with different variational optimization strategies, including COBYLA, Adam, and L-BFGS-B in the presence of finite sampling noise. These resource estimates provide realistic expectations for embedding this quantum solver into DMFT workflows on current quantum hardware.

The remainder of this paper is organized as follows. In Section~\ref{sec:methods}, we describe the methods used to calculate the continued fraction representation of the impurity Green's function, present the ansatz architecture employed to approximate the ground and excited states of the AIM, and introduce the tested optimization routines. Section~\ref{sec:results} reports our numerical results, including variational ground-state preparation, the fidelity and cost of optimization, moment-based energy corrections, and Green's function reconstruction. We conclude in Section~\ref{sec:outlook} with a summary of our findings and a discussion of future directions for integrating quantum impurity solvers within dynamical mean-field theory workflows.

\section{Methods}
\label{sec:methods}

In this section, we outline the workflow developed to compute the impurity Green's function. We first define the Hamiltonian used to model the interacting impurity coupled to a non-interacting bath, followed by a description of the ansatz used to approximate its ground state via VQE and the methods used to optimize it. To improve the variational energy estimates, we discuss how to incorporate leading corrections derived from a cumulant expansion based on low-order Hamiltonian moments. Finally, we use our ans\"atze to approximate particle and hole excitations and the parameters of the optimized ground-state circuit, and describe the Krylov subspace procedure used to construct the continued-fraction representation of the impurity Green's function.

\subsection{The impurity model}
\label{sec:AIM}

This work focuses on single-impurity AIMs defined by the Hamiltonian~\cite{anderson_pw_1961}
\begin{gather}
    H_{\rm AIM} = H_{\rm imp} + H_{\rm bath} + H_{\rm coup} ,\nonumber\\
    H_{\rm imp} = \sum_\sigma \epsilon_0 n_{0\sigma} + U n_{0\uparrow} n_{0\downarrow} ,\nonumber\\
    H_{\rm bath} = \sum_{k=1}^{N_b} \sum_\sigma \epsilon_k n_{k\sigma} ,\nonumber\\
    H_{\rm coup} = \sum_{k=1}^{N_b} \sum_\sigma V_k \left( c_{0\sigma}^\dagger c_{k\sigma} + c_{0\sigma} c_{k\sigma}^\dagger \right) ,
\label{eq:aimHam}
\end{gather}
where $\epsilon_0$ and $\epsilon_k$ are the single-particle energies of the impurity and the $k$-th bath site, respectively, $U$ the Hubbard interaction, $V_k$ is the hybridization between the impurity and the $k$-th bath sites, $N_b\in\{1,3,5,7\}$ is the number of bath sites, and $\sigma\in\{\uparrow,\downarrow\}$ is the projection of the electron's spin. The operators $c_{i\sigma}^\dagger$ and $c_{i\sigma}$ create and annihilate an electron with spin projection $\sigma$ in site $i$. The number operators $n_{i\sigma}$ count the number of $\sigma$-electrons in orbital $i$.

In this work, we impose particle–hole symmetry on the Hamiltonian by setting $\epsilon_0=-U/2$, ensuring the system is half-filled, and its ground state has spin projection $S_z=0$ in the absence of an external magnetic field, and set $V_k=V$ for all $k$.

These impurity models form the foundation of DMFT, which maps the original interacting lattice problem onto a self-consistently defined quantum impurity coupled to a non-interacting bath. This mapping retains all local interaction effects while approximating non-local correlations. The interaction between the impurity and the bath is described by the hybridization function, which reflects the dynamical coupling to the environment and must be updated iteratively through the DMFT loop. In practical implementations, this continuous bath is discretized into a finite number of orbitals, making the bath size a crucial control parameter. As the bath size increases, the model captures finer dynamical features, but also becomes increasingly challenging to solve. These challenges motivate the development of quantum impurity solvers~\cite{sakurai_2022, sakurai_2024} capable of efficiently handling large Hilbert spaces and providing direct access to real-frequency observables.

Solving the impurity problem in DMFT involves computing the impurity Green's function~\cite{kotliar_g_2006, georges_a_1996}, which as a function of the energy $\omega$ takes the from
\begin{align}
    G_{ij}(\omega) =& \braket{\psi_0|c_i \frac{1}{\omega-H_{\rm AIM}+E_0} c_j^\dagger|\psi_0} \nonumber\\
    +& \braket{\psi_0|c_j^\dagger \frac{1}{\omega+H_{\rm AIM}-E_0} c_i|\psi_0} .
    \label{eq:impGF}
\end{align}
Here, the ground state $\psi_0$ of the AIM Hamiltonian~\eqref{eq:aimHam} solves the Sch\"odinger equation
\begin{equation}
    H_{\rm AIM}\ket{\psi_0} = E_0\ket{\psi_0} .
    \label{eq:aimGS}
\end{equation}
Therefore, computing the impurity Green's function involves finding the ground state of the AIM~\eqref{eq:aimHam}, calculating its energy, and evaluating the matrix elements on the right-hand side of Eq.~\eqref{eq:impGF}.

\subsection{Variational ground state approximation}
\label{sec:GS_methods}

Our quantum solver employs VQE to optimize a PQC minimizing the variational energy
\begin{equation}
   E_{\rm VQE}(\boldsymbol{\theta}) = \underset{\boldsymbol{\theta}}{\rm min} \braket{\psi(\boldsymbol{\theta})|H_{\rm AIM}|\psi(\boldsymbol{\theta})} ,
\label{eq:ev}
\end{equation}
to find an approximation $\ket{\boldsymbol{\theta}}$ to the model's ground state $\ket{\psi_0}$. To this end, the Hamiltonian~\eqref{eq:aimHam} is recast in terms of spin operators using the Jordan–Wigner transformation~\cite{jordan_wigner_1928, bravyi_2002},
\begin{align}
    c_i^\dagger =& \prod_{j=0}^{i-1}Z_j \otimes \frac{1}{2}(X_i - i Y_i) \otimes \cdots \nonumber \\
    c_i =& \prod_{j=0}^{i-1}Z_j \otimes \frac{1}{2}(X_i + i Y_i) \otimes \cdots ,
\label{eq:jw}
\end{align}
where $X_i$, $Y_i$ and $Z_i$ are Pauli matrices acting on the space of qubit $i$, which encodes the occupation of a single spin-orbital in the model.

\begin{figure}[t]
    \centering
    \includegraphics[width=0.95\linewidth]{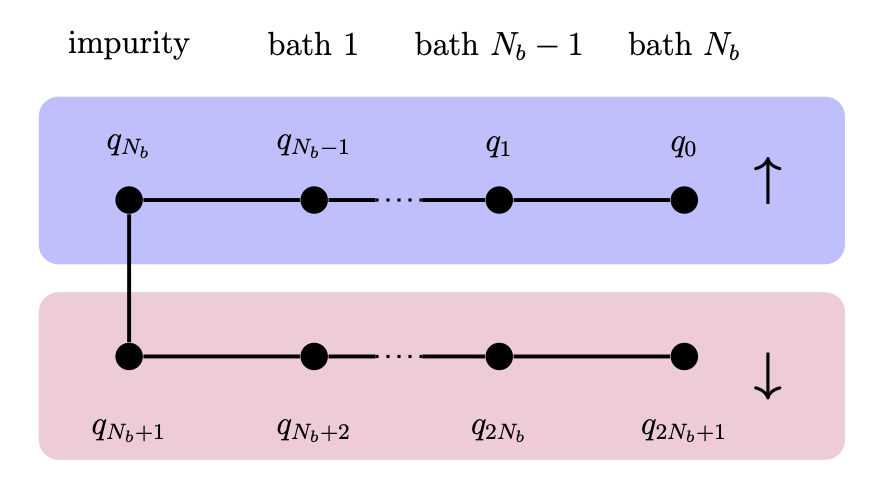}
    \caption{Spin-block ordering used to map the model’s spin-up and down sites onto a linear chain of qubits.}
    \label{fig:order}
\end{figure}

Using the orbital-to-qubit mapping depicted in Fig.~\ref{fig:order} to assign qubits along a linear chain to the model's spin-orbitals leads to the expression
\begin{align}
    H_{\rm AIM} =& C - \sum_{k=1}^{N_b} \frac{\epsilon_k}{2} (Z_{N_b-k} + Z_{N_b+1+k}) + \frac{U}{4} Z_{N_b}Z_{N_b+1} \nonumber\\
    +& \frac{V}{2} \sum_{k=1}^{N_b} X_{N_b-k} \left( \prod_{j=N_b+1-k}^{N_b-1} Z_j \right) X_{N_b} \nonumber\\
    +& \frac{V}{2} \sum_{k=1}^{N_b} Y_{N_b-k} \left( \prod_{j=N_b+1-k}^{N_b-1} Z_j \right) Y_{N_b} \nonumber\\
    +& \frac{V}{2} \sum_{k=1}^{N_b} X_{N_b+1} \left( \prod_{j=N_b+2}^{N_b+k} Z_j \right) X_{N_b+1+k} \nonumber\\
    +& \frac{V}{2} \sum_{k=1}^{N_b} Y_{N_b+1} \left( \prod_{j=N_b+2}^{N_b+k} Z_j \right) Y_{N_b+1+k} ,
\end{align}
where $C=-U/4+\sum_{k=1}^{N_b} \epsilon_k$.

The expectation value~\eqref{eq:ev} is calculated adding the expectation values of each term in the Pauli-strings decomposition of the Hamiltonian, which can be estimated on a quantum chip by measuring the state $\psi(\boldsymbol{\theta})$ in the $N_{\rm group}$ bases corresponding to the commuting groups into which the Pauli strings are organized. The absolute error associated with the finite sampling of the state is estimated as
\begin{equation}
    \varepsilon \sim \mathcal{O}\left(\sqrt{N_{\rm Pauli}/N_{\rm shots}}\right) ,
\label{eq:samperr}
\end{equation}
where $N_{\rm Pauli}$ and $N_{\rm shots}$ denote the number of Pauli strings defining the Hamiltonian and the number of measurements taken on each basis~\cite{crawford_2021, huang_2021}.

\begin{figure*}[ht]
    \centering
    \includegraphics[height=0.16\linewidth, trim={0 0 0 0}, clip]{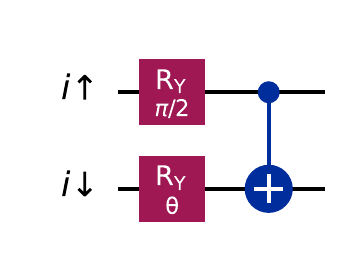}
    \includegraphics[height=0.16\linewidth, trim={0 0 0 0}, clip]{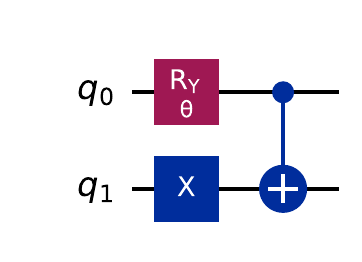}
    \includegraphics[height=0.16\linewidth, trim={0 0 0 0}, clip]{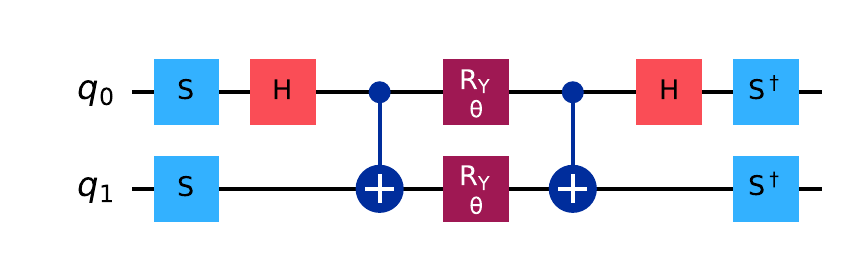}
    \includegraphics[height=0.29\linewidth, trim={0 0 0 0}, clip]{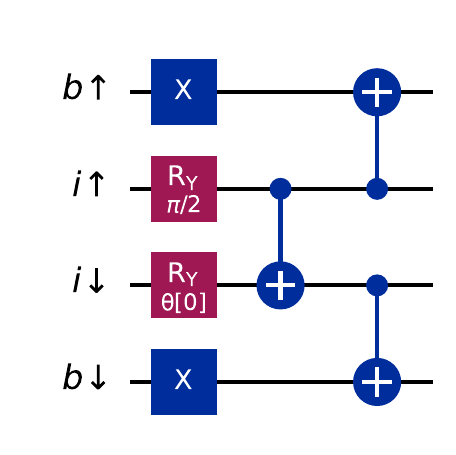}
    \includegraphics[height=0.285\linewidth, trim={0 0 0 0}, clip]{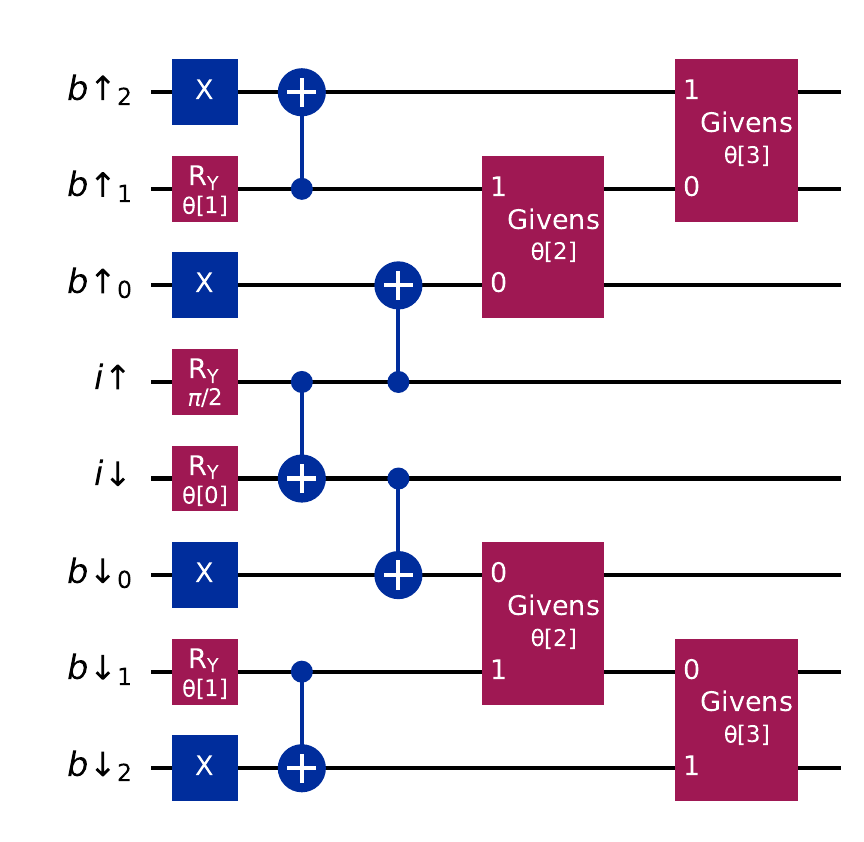}
    \includegraphics[height=0.285\linewidth, trim={0 0 0 0}, clip]{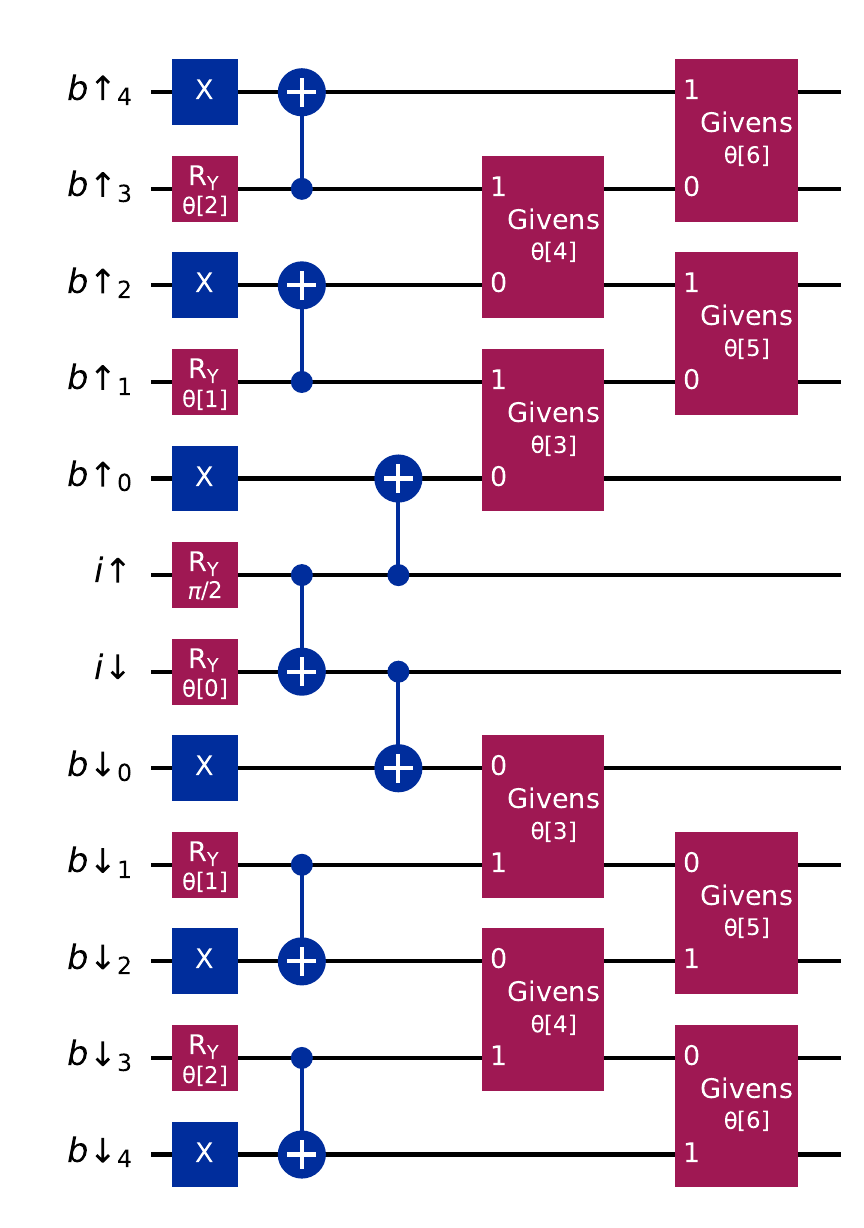}
    \includegraphics[height=0.285\linewidth, trim={0 0 0 0}, clip]{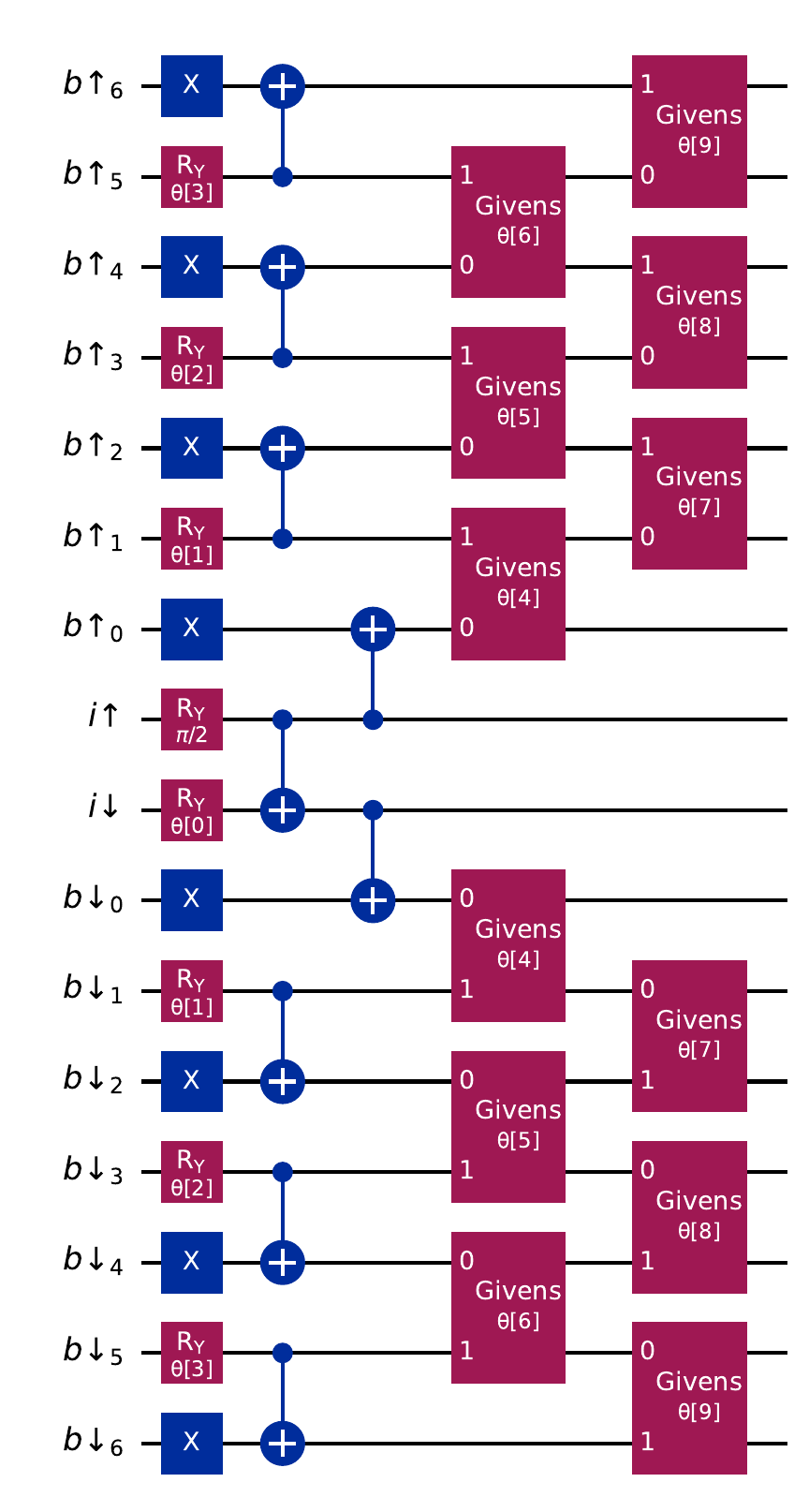}
    \caption{Building blocks used to construct symmetry-preserving ans\"atze (top), and ans\"atze for the ground states of particle-hole symmetric AIMs with one, three, five, and seven bath sites (bottom). These ans\"atze first prepare a spin-exchange symmetric impurity state (with the top left circuit), half-fill the rest of the orbitals (with the top center one) and implements Givens rotation (via the top right circuit) to prepare states with the proper number of electrons.}
    \label{fig:gs_ansatz}
\end{figure*}

The symmetries of the ground state arising from that imposed on the Hamiltonian allow us to propose a highly constrained PQC as an ansatz for the ground state. The first component of this PQC, shown at the top left of Fig.~\ref{fig:gs_ansatz}, acts on the impurity qubits initializing them to
\begin{align}
    \ket{\psi(\theta)} = \sqrt{\frac{1}{2}} \bigg\{ &\cos\left( \frac{\theta}{2}\right) \left( \ket{00} + \ket{11} \right) \nonumber\\
    + &\sin\left( \frac{\theta}{2}\right) \left( \ket{01} + \ket{10} \right) \bigg\} ,
\end{align}
a state that is symmetric under spin exchange.

Next, the half-filling gate at the top center of Fig.~\ref{fig:gs_ansatz}, which ``creates'' an electron in a spin-orbital pair, is used to prepare an initial state with the proper number of electrons. The most general state with this electron number can be prepared by acting the right number of Givens rotation gates, shown at the top right of Fig.~\ref{fig:gs_ansatz}, on the initial state~\cite{gard_bt_2020}. In this work, however, we restrict the number of Givens rotations to those that can be implemented within two layers after the initial state preparation (see, for example, the three-bath ansatz at the bottom right of Fig.~\ref{fig:gs_ansatz}), thus constraining the PQC's local depth in an attempt to mitigate barren plateaus~\cite{zhang_hk_2024}. More expressive symmetry-preserving ans\"atze including additional entangling structures or repetitions of the variational layer have been proposed and used to accurately prepare AIM ground states~\cite{jones_eb_2025}. Here, we adopt a more constrained variant to maintain shallow depths that could help with its optimization on present-day noisy hardware, and allow for the preparation of particle and hole excitations from the same parameter set, eliminating additional optimization costs.

To enforce spin-exchange symmetry, we apply identical unitaries to the top and bottom registers. The state prepared by the PQC is
\begin{align}
    \ket{\psi(\boldsymbol{\theta})} =& \sum_a c_{a00a}(\boldsymbol{\theta}) \ket{a}\otimes\ket{00}\otimes\ket{a} \nonumber\\
    +& \sum_a c_{a11a}(\boldsymbol{\theta}) \ket{a}\otimes\ket{11}\otimes\ket{a} \nonumber\\
    +& \sum_a\sum_b c_{a01b}(\boldsymbol{\theta}) \ket{a}\otimes\ket{01}\otimes\ket{b} \nonumber\\
    +& \sum_a\sum_b c_{b10a}(\boldsymbol{\theta}) \ket{b}\otimes\ket{01}\otimes\ket{a} ,
\end{align}
where $a,b$ are bitstrings representing basis states of the bath registers, and $c_{aijb}=c_{bjia}$. In particular, for the one-bath case we have
\begin{align}
    \ket{\psi(\theta_0)} = \sqrt{\frac{1}{2}} \bigg\{ &\cos\left(\frac{\theta_0}{2}\right) \left( \ket{1001} + \ket{0110} \right) \nonumber\\
    +& \sin\left(\frac{\theta_0}{2}\right) \left( \ket{1010} + \ket{0101} \right) \bigg\} .
\label{eq:1bgs}
\end{align}

The ans\"atze for the ground states of single-impurity AIMs with one, three, five and seven bath sites are shown at the bottom of Fig.~\ref{fig:gs_ansatz}. Notice that the ansatz for the one-bath case is embedded within those for the ground states of AIMs with more than one bath site. 

\subsection{Optimization methods}
\label{sec:opt_methods}

The estimation of the variational energy depends on finding the parameters that best approximate the ground state, making the choice of optimizer critical for the efficient and accurate computation of this observable. Different optimization strategies navigate the parameter landscape in distinct ways, each presenting trade-offs between convergence speed, sensitivity to sampling noise, and computational cost. To identify the most effective approach, we examined three representative optimization strategies employing gradient-free, first-order, and quasi-second-order methods.

The constrained optimization by linear approximation (COBYLA) gradient-free method~\cite{Powell1994} constructs a local linear model of the objective function by evaluating it at a dynamical number of parameter points located within a hypersphere of radius $\rho_0$, set in this work to $\pi$, centered at the initial point. The parameters that minimize this model are then chosen as a new center, and the procedure is repeated for a new radius, $\rho_1<\rho_0$. Hence, gradient-free optimizers has the lowest measurement cost per evaluation, requiring only the $N_{\rm group}$ circuit measurements needed to estimate the energy.

Adam was selected as a representative first-order optimizer~\cite{kingma_2017}. Gradient-based methods update the parameters along the direction of the negative gradient, estimated on a quantum computer using the parameter-shift rule,
\begin{equation}
    \frac{\partial E_{\rm VQE}}{\partial \theta_i}\bigg|_{\boldsymbol{\theta}} = \frac{1}{2} \left[ E_{\rm VQE}(\boldsymbol{\theta}_i^+) - E_{\rm VQE}(\boldsymbol{\theta}_i^-) \right] ,
\end{equation}
with $\boldsymbol{\theta}_i^\pm$ a set of parameters in which the $i$-th component of $\boldsymbol{\theta}$ is set to $\theta_i\pm\pi/2$. Although the number of measurements per evaluation, $N_{\rm group}(1 + 2N_{\rm params})$, is substantially higher than that for gradient-free methods, first-derivative information may help these optimizers converge the VQE in fewer overall evaluations. Our Adam optimizations used an initial learning rate $\eta=1$, standard hyperparameters $\{\beta_1,\beta_2,\epsilon\}=\{0.9, 0.999,10^{-8}\}$, and initialized all momentum terms to zero.

For comparison, we also tested the quasi-second-order limited-memory Broyden-Fletcher-Goldfarb-Shanno (L-BFGS-B) algorithm~\cite{Byrd1995}, which approximates the inverse Hessian from a limited number of past gradient estimates. In this context, the number of measurements per evaluation is that of a first-order gradient-based method.
  
Convergence for all optimizers was defined as either an absolute estimate change $|E_{\rm VQE}^{(n)}-E_{\rm VQE}^{(n-1)}|<10^{-6}$, with $E_{\rm VQE}^{(n)}$ the energy estimated at iteration $n$, or a gradient norm $||\nabla_{\boldsymbol{\theta}}E_{\rm VQE}^{(n)}||<10^{-9}$.

\subsection{Quantum computed moments (QCM) approach} 

A key limitation of the VQE is that the extent to which the state $\psi(\boldsymbol{\theta})$ and the estimated energy $E_{\rm VQE}$ approximate the true ground state and its energy strongly depends on the expressive power of the chosen ansatz and the convergence of the optimization routine. The latter is severely affected by errors in the expectation value and the presence of barren plateaus.

The QCM approach proposed in Ref.~\cite{vallury_2020_quantum} addresses this limitation by estimating a correction to the variational energy from higher-order Hamiltonian moments $\mu_m=\langle \psi(\boldsymbol{\theta})| H^m | \psi(\boldsymbol{\theta}) \rangle$. In terms of a set of cumulants $c_m$, defined as
\begin{equation}\label{eq:c_n}
    c_m = \mu_m - \sum\limits_{p=0}^{m-2} \binom{m-1}{p}\, c_{p+1}\, \mu_{m-1-p} ,
\end{equation}
the Lanczos coefficients~\cite{lanczos_1950_an} defining the tridiagonal Hamiltonian in the Krylov basis built on $\psi(\boldsymbol{\theta})$,
\begin{equation}
    H_{\rm AIM} = \left( \begin{array}{c c c c}
         a_0 & b_1 & 0 & \ldots \\
         b_1 & a_1 & b_2 \\
         0 & b_2 & a_3 \\
         \vdots & & & \ddots
    \end{array} \right) ,
\end{equation}
can be written as expansions in powers of $1/N$~\cite{hollenberg_1993_plaquette},
\begin{align}
    a_\ell =& c_1 + \ell \left(\frac{c_3}{c_2}\right) \frac{1}{N} + \mathcal{O}(N^{-2}) \nonumber\\
    b_\ell^2 =& \ell c_2 + \frac{1}{2} \ell(\ell-1) \left(\frac{c_2c_4-c_3^2}{2c_2^2}\right) \frac{1}{N} + \mathcal{O}(N^{-2}) ,
\label{eq:lanczos_expansion}
\end{align}
with $N$ the size of the system. For large $\ell$ and $N$, these expansions become series in powers of $z=\ell/N$, in terms of which the ground-state energy can be calculated as the infimum of $a(z)-2b(z)$ for $z>0$~\cite{hollenberg_1996_analytic, hollenberg_1994_general}. At first order in $z$, this infimum takes the approximate form
\begin{equation}\label{eq:E_inf}
    E_{\rm INF} \approx c_1 - \frac{c_{2}^{2}}{c_{3}^{2}-c_{2}\,c_4}\, \left( \sqrt{3c_{3}^{2} - 2c_{2}\,c_{4}} - c_{3}\right) .
\end{equation}

Notice that this approximation, although no longer a strict upper bound, incorporates corrections from higher-order Hamiltonian moments in its second term, yielding a more accurate estimate of the true ground-state energy than the variational result alone. From a computational perspective, this approach shifts part of the problem complexity away from deep, highly expressible circuits to the measurement of higher-order observables.

Corrections to the ground-state energy are particularly relevant within the context of DMFT, as the accurate determination of this quantity is important to stabilize the self-consistency loop. By incorporating higher Hamiltonian moments, one can potentially mitigate errors due to shallow ansatz, sampling noise, and barren plateaus.

\subsection{Green's function as continued fractions}\label{sec:lanczos_methods}

In order to extract physically meaningful observables from the quantum-computed ground and excited states of the AIM, we reconstruct the impurity Green's function in the real-frequency domain using a Krylov subspace-based continued-fraction expansion.

Since the model lacks coupling between spin-up and spin-down orbitals, the off-diagonal elements of the impurity Green’s function~\eqref{eq:impGF} vanish identically. The diagonal matrix elements can be expressed as continued fractions~\cite{haydock_1972}
\begin{align}\label{eq:G_ii}
    G_{ii}(\omega) =& \frac{1}{\omega+E_0-a_0^p-\frac{(b_1^p)^2}{\omega+E_0-a_1^p-\cdots}} \nonumber\\
    +& \frac{1}{\omega-E_0+a_0^h-\frac{(b_1^h)^2}{\omega-E_0+a_1^h-\cdots}} ,
\end{align}
where $a_n^p$ and $b_n^p$ are the Lanczos coefficients of $H_{\rm AIM}$ on the Krylov basis built on top of $c_i^\dagger\ket{\psi_0}$, and $a_n^h$ and $b_n^h$ the corresponding coefficients on the basis built on top of $c_i\ket{\psi_0}$.

\begin{figure}[ht]
    \centering
    \includegraphics[height=0.425\linewidth]{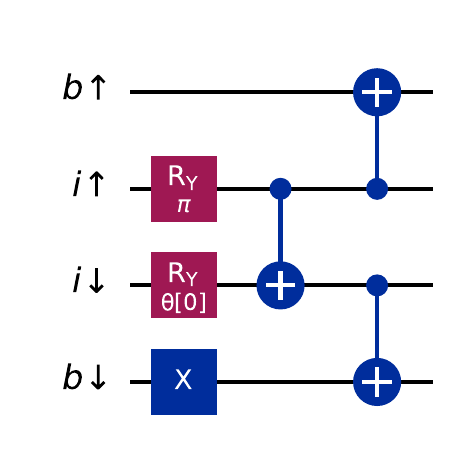} \includegraphics[height=0.425\linewidth]{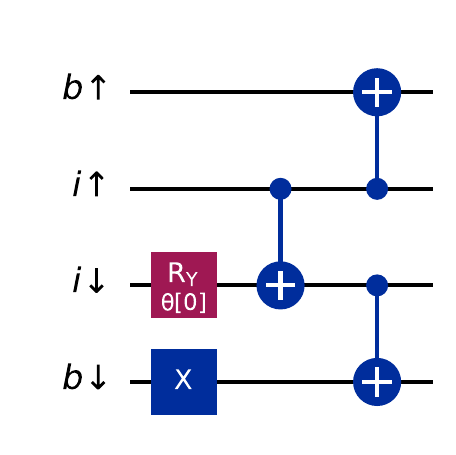}
    \includegraphics[height=0.425\linewidth]{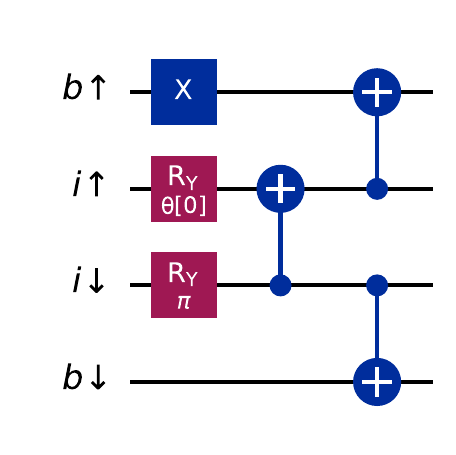} \includegraphics[height=0.425\linewidth]{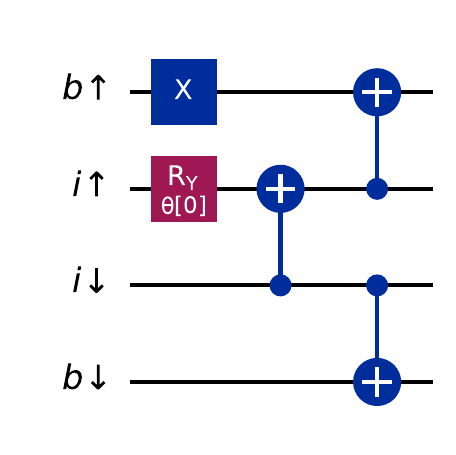}
    \caption{Ans\"atze for particle (left) and hole (right) excitations of the groud state of AIMs with one bath site, preparing the states $\mathscr{N}c_{i\sigma}^\dagger\ket{\rm GS}$ and $\mathscr{N}c_{i\sigma}\ket{\rm GS}$ with spin projections $\sigma=\uparrow$ (top) and $\sigma=\downarrow$ (bottom).}
    \label{fig:ansatz_excited}
\end{figure}

The ans\"atze in Fig.~\ref{fig:ansatz_excited} can be used to prepare normalized versions of these particle and hole excitations.
The PQCs at the top of the figure prepare the states
\begin{align}
    \ket{\phi(\theta_0)} =& \cos\left(\frac{\theta_0}{2}\right) \ket{0111} + \sin\left(\frac{\theta_0}{2}\right) \ket{1011} \nonumber\\
    \ket{\chi(\theta_0)} =& \cos\left(\frac{\theta_0}{2}\right) \ket{1000} + \sin\left(\frac{\theta_0}{2}\right) \ket{0100} ,
\end{align}
with one and three electrons. For the parameter $\pi-\theta_0$,
\begin{align}
    \ket{\phi(\pi-\theta_0)} 
    =& \mathscr{N} c_{i\uparrow}^\dagger\ket{\psi(\theta_0)} \nonumber\\
    \ket{\chi(\pi-\theta_0)} 
    =& \mathscr{N} c_{i\uparrow}\ket{\psi(\theta_0)} ,
\end{align}
with $\ket{\psi(\theta_0)}$ the one-bath state in Eq.~\eqref{eq:1bgs}, and $\mathscr{N}$ a normalization constant. Similarly, the PQCs at the bottom of Fig.~\ref{fig:ansatz_excited} prepare states proportional to $c_{i\downarrow}^\dagger\ket{\psi(\theta_0)}$ and $c_{i\downarrow}\ket{\psi(\theta_0)}$.

Substituting the one-bath ground-state ansatz embedded in the ans\"atze for the ground states of AIMs with more than one bath site, with the PQCs shown in Fig.~\ref{fig:ansatz_excited}, and replacing $\boldsymbol{\theta}$ with a new parameter set in which $\theta_0$ is replaced by $\pi-\theta_0$, yields the corresponding particle and hole excitations $\mathscr{N}c_{i\sigma}^\dagger\ket{\psi(\boldsymbol{\theta})}$ and $\mathscr{N}c_{i\sigma}\ket{\psi(\boldsymbol{\theta})}$, with $\sigma\in\{\uparrow,\downarrow\}$.

Using the ans\"atze for the particle and hole excitations we compute the Lanczos coefficients in Eq.~\eqref{eq:G_ii} as~\cite{witte_ns_1994, jones_wb_1980}
\begin{gather}
a_n = \frac{\Delta'_{n-2}\Delta_n}{\Delta'_{n-1}\Delta{n-1}} + \frac{\Delta'_n\Delta_n-1}{\Delta'_{n-1}\Delta_n} \nonumber\\
b_n^2 = \frac{\Delta_n\Delta_{n-2}}{\Delta_{n-1}^2} ,
\label{eq:hankel_lanczos}
\end{gather}
with
\begin{gather}
    \Delta_n = \left| \begin{array}{cccc}
        \mu_0 & \mu_1 & \ldots & \mu_n \\
        \mu_1 & \mu_2 \\
        \vdots & & \ddots \\
        \mu_n & & & \mu_{2n}
    \end{array}\right| \ \Delta'_n = \left| \begin{array}{cccc}
        \mu_1 & \mu_2 & \ldots & \mu_{n+1} \\
        \mu_2 & \mu_3 \\
        \vdots & & \ddots \\
        \mu_{n+1} & & \ldots & \mu_{2n+1}
    \end{array}\right|
\end{gather}
Hankel determinants. Since the number of Pauli strings, and therefore the number of quantum circuits, involved in the estimation of $\mu_m$ increases rapidly with $m$, we estimate Lanczos coefficients up to order $n=2$, corresponding to the computaion of moments up to $\mu_5$.

Preparing particle and hole excitations enables the direct evaluation of the moments $\mu_m$, rather than the indirect ground-state-based approach in which powers of the Hamiltonian are dressed with creation and annihilation operators~\cite{greenediniz_g_2024}. This construction reduces the number of Pauli strings entering the moments' estimation and the number of shots required to achieve a target absolute error $\epsilon$.

\subsection{Classical reference calculations}
\label{sec:classical-calculation}

To validate our quantum ansatz, we also perform classical calculations of the Green's function. For the model sizes investigated in the present paper, exact diagonalization of the Hamiltonian is easily accomplished without the need for high-performance computing resources. We follow the procedure outlined in section~\ref{sec:lanczos_methods} above to represent the Green's function as a continued fraction with its coefficients being determined using Lanczos iteration. For this, we first construct a basis for the ground state calculations, which consists of all sets of creation operators that preserve the total charge and spin for the half-filled system considered here, i.e. a total of $2 \binom{N_b+1}{(N_b+1)/2}$ of basis states. The Hamiltonian for the AIM can be represented as a sparse matrix and we first find the lowest eigenvalue state $\ket{\psi_0}$ with the ground state energy $E_0$. Using this, we can apply creation or annihilation operators to this vector and follow Eq.~\eqref{eq:impGF} to calculate the many-body Green's function using the continued fraction in Eq.~\eqref{eq:G_ii}, where the coefficients can be obtained using Lanczos iteration. \cite{haydock_1972,haydock_1980}

\section{Results}
\label{sec:results}

The results presented in this section, used to assess the performance of the proposed quantum solver, were obtained from classical simulations of the quantum subroutines involved in the computation of the impurity Green's function~\eqref{eq:impGF}. These simulations solved the AIM described in Section~\ref{sec:AIM}, with Hubbard interactions $U\in\{2,4,6,8\}$. To account for the potential for the optimizer to become trapped in local minima, the VQE results for AIMs with one, three, and five bath sites reported here are averaged over simulations initiated from ten distinct starting points. Moreover, each optimization was repeated five times to mitigate statistical errors arising from the finite number of shots used to evaluate the variational estimates. For the more computationally demanding seven-bath case, simulations were performed using a single initialization, with two optimization repeats. The scalability of the solver with respect to the number of bath sites is assessed by averaging results over the different values of $U$, initializations, and optimization repetitions. As discussed in Section~\ref{sec:opt_methods}, the ansatz was optimized using a gradient-free method (COBYLA), a first-order gradient-based method (Adam), and a quasi-second-order gradient-based method (L-BFGS-B) to compare the performance of different optimization strategies.

\subsection{Ground-state optimization}\label{sec:GS_results}

\begin{figure}[b]
    \centering
    \includegraphics[width=0.95\linewidth]{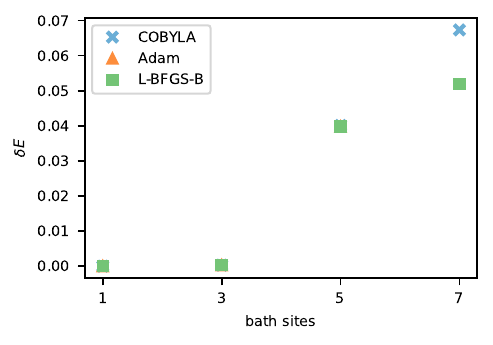}
    \caption{Best absolute relative error between true ground-state energies and variational estimates for them obtained from noiseless simulations. For models with seven bath sites, the proposed ansatz can yield energy estimates accurate to within about five percent of the true ground-state energy.}
\label{fig:ansatz_performance}
\end{figure}

The ansatz proposed for the ground state does not span the full $n$-qubit Hilbert space and therefore may struggle to approximate the true ground state of the AIM as the number of bath sites increases. To assess the performance of our ansatz, we compared ground-state energies obtained via classical diagonalization with those produced by noiseless VQE simulations. Figure~\ref{fig:ansatz_performance} shows the absolute relative errors $\delta E\equiv {\rm abs}([E_{\rm ideal}-E_{\rm exact}]/E_{\rm exact})$ produced by these simulations.
For AIMs with five and seven bath sites, the ansatz was capable of yielding variational energies that deviate from the exact ground-state energy by approximately five percent.

\begin{figure}[t]
    \centering
    \includegraphics[width=0.95\linewidth]{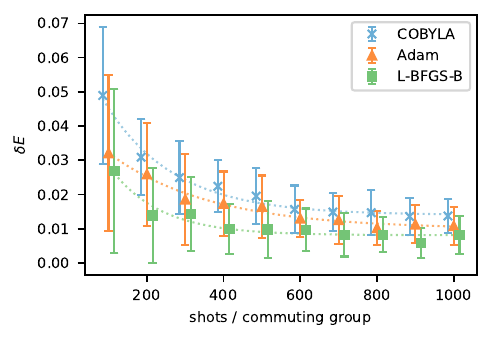}
    \includegraphics[width=0.95\linewidth]{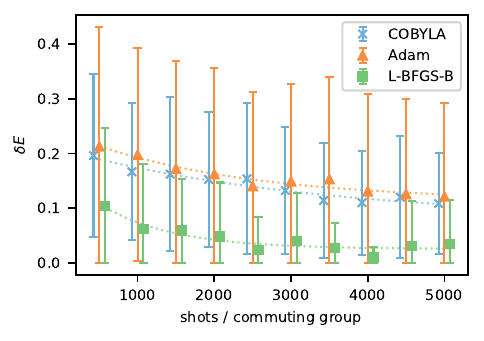}
    \caption{Averaged absolute relative error between converged variational energies in the presence and absence of sampling noise. Simulations solving AIMs with on-site energy $U=2$ and one (top) or three (bottom) bath sites show a decrease of the relative error with the number of shots. Means and standard deviations result from simulations initiated from ten distinct starting points, each repeated five times to account for optimization difference due to sampling noise.}
    \label{fig:samp_error}
\end{figure}

In practice, the variational energy is estimated from a finite number of measurements, causing sampling noise that may impede convergence to the best achievable approximation to the ground state. To quantify this effect, we performed VQE simulations with a varying number of shots for AIMs with one and three bath sites and Hubbard interaction $U=2$. The resulting absolute relative errors, $\delta E\equiv{\rm abs}([E_{\rm samp}-E_{\rm ideal}]/E_{\rm ideal})$, are shown in Fig.~\ref{fig:samp_error}.

These values show that for all three optimizers, the relative energy error decreases as the number of shots per commuting group increases. The rate and stability of this improvement depend strongly on whether gradients are used in the optimization. Across the range of shots, COBYLA exhibits large errors, indicating that sampling noise can cause the gradient-free search to get trapped in local minima or terminate early. In contrast, gradient-based methods can achieve higher accuracy, but doing so requires information about second derivatives. For the best performing optimization method (L-BFGS-B), relative differences of roughly two percent are achieved in the one- and three-bath cases from 100 and 2500 measurements per quantum circuit, respectively. This behavior highlights that reliable gradient information and accurate Hessian approximations are crucial to avoid noise-induced convergence to suboptimal states.

Using relation~\eqref{eq:samperr}, we estimate that absolute errors $\epsilon\sim 0.1$ on each energy estimate can be achieved with 700, 2500, 5000 and 7500 shots per commuting group for AIMs with one, three, five, and seven bath sites, respectively.

\subsection{Optimization performance, fidelity, and cost}\label{sec:performance_results}

\begin{figure}[b]
    \centering
    \includegraphics[width=0.95\linewidth]{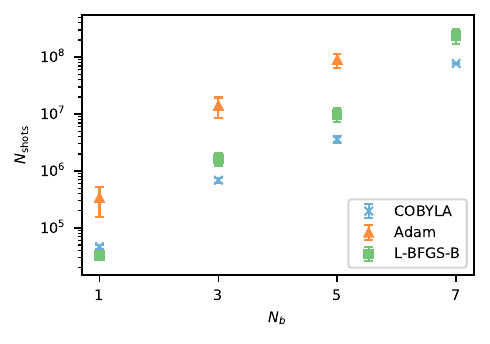}
    \caption{Comparison between averaged total number of shots required by different optimization methods to converge VQEs solving single-impurity AIMs with $N_b\in\{1,3,5,7\}$.}
    \label{fig:vqe_num_qcs}
\end{figure}

We use the number of shots required to converge the VQE as a metric of the subroutine's quantum cost. Figure~\ref{fig:vqe_num_qcs} shows the scaling of this cost with the number of bath sites for the three different optimization methods. 
The costs presented there, which are averages over all runs that solved AIMs with the corresponding number of bath sites, highlight the differences in computational expense between the tested optimization methods.

The COBYLA optimizer measures the lowest number of shots per evaluation and thus exhibits the best scaling with $N_b$. Unfortunately, without the computation of a gradient pushing in the direction of maximum change, the objective function tends to get stuck in local minima, as reflected by the larger $\delta E$ values for this method in Fig.~\ref{fig:samp_error}. The gradient-based Adam and L-BFGS-B optimizers substantially increase the number of circuit measurements per evaluation, leading to worse scaling than that of COBYLA. Furthermore, results in Fig.~\ref{fig:samp_error} from runs using the Adam optimizer suggest that sampling errors in the gradient estimate can push the optimization towards local minima, yielding larger $\delta E$ values. This effect seems to be mitigated via the estimation of the inverse Hessian carried on by the L-BFGS-B optimizer.

These cost differences are considered together with the accuracy results discussed in Sec.~\ref{sec:GS_results}. In ideal simulations, the proposed ansatz produces ground-state energies that differ from the exact values by at most a few percent for $N_b \leq 7$, as shown in Fig.~\ref{fig:ansatz_performance}, indicating that the ansatz retains a reasonable ground-state fidelity across the range of bath sizes studied in this work. In the presence of sampling noise, COBYLA is more prone to being trapped in shallow local minima, especially at low shot counts. This leads to larger energy errors despite its lower circuit cost. The gradient-based optimizers make more efficient use of the measurement statistics. For a fixed target accuracy, they achieve smaller relative energy errors at moderate shot numbers, with L-BFGS-B providing the most favorable balance between accuracy and quantum cost.

These results highlight a trade-off between circuit depth, optimizer robustness, and measurement overhead. Gradient-free methods minimize the raw number of circuits but can incur a significant accuracy penalty, particularly when measurements are noisy. Gradient-based methods require more circuits but deliver higher-fidelity approximations to the impurity ground state.

\subsection{Energy correction estimates}\label{sec:qcm_results}

As discussed in Sec.~\ref{sec:GS_methods}, the variational energy can be refined by incorporating higher-order Hamiltonian moments measured in the optimal state. Starting from the moments $\mu_n$ up to $n=4$, we used the approximation to the infimum~\eqref{eq:E_inf}. This approximation corrects the raw variational estimate $c_1$ with fluctuations encoded in $c_2$, $c_3$, and $c_4$, which capture increasingly nontrivial dynamical correlations beyond those described by the shallow ansatz alone.

For consistency, the number of shots used to estimate $\mu_4$ is increased to achieve an absolute error of $\epsilon\sim 10^{-1}$. These numbers are listed in Table~\ref{tab:shots}. From those, the total number of shots required to compute the correction to the variational energy are approximately $4.0\times10^3$, $6.6\times10^6$, and $2.8\times10^8$ for the one, three, and five-bath cases.
\begin{table}[t]
    \centering
    \begin{tabular*}{0.95\columnwidth}{@{\extracolsep{\fill}} c c c c c c c c @{}}
        \hline\hline
        & \multicolumn{3}{c}{$N_{\rm Pauli}$} & \multicolumn{2}{c}{$N_{\rm group}$} & \multicolumn{2}{c}{$N_{\rm shots}$} \\
        \cline{2-4} 
        \cline{5-6}
        \cline{7-8}
        $N_b$ & $H$ & $H^4$ & $H^5$ & $H^4$ & $H^5$ & $\mu_4$ & $\mu_5$ \\
        \hline
        1 & 6 & 23 & 24 & 2 & 2 & 2K & 2K \\
        3 & 18 & 1190 & 2146 & 60 & 63 & 110K & 185K \\
        5 & 30 & 10995 & 33672 & 278 & 597 & 1M & 3M \\
        7 & 42 & 47000 & 212222 & 859 & 2143 & 4M & 18M \\
        \hline\hline
    \end{tabular*}
    \caption{Number of Pauli strings in $H^n$ up to $n=4$, number of commuting groups in $H^n$ with $n\in\{4,5\}$, and number of shots per group required to achieve absolute errors $\epsilon\sim 10^{-1}$ on estimates for $\mu_n$ with $n\in\{4,5\}$.}
    \label{tab:shots}
\end{table}

\begin{figure*}[ht]
    \centering
    \includegraphics[width=0.95\linewidth]{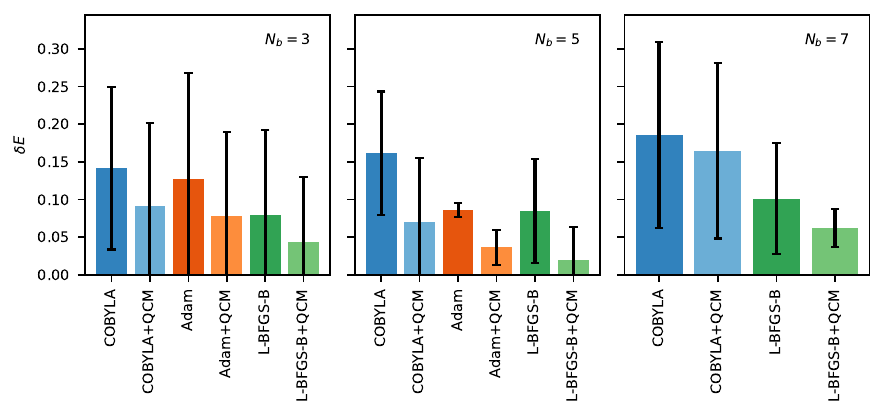}
    \caption{Average absolute relative error $\delta E$ between the exact ground-state energy and its variational estimate for AIMs with three (left), five (center), and seven (right) bath sites. Bars compare the variational energies obtained with the COBYLA, Adam, and L-BFGS-B methods to the corresponding QCM-corrected estimates (``+QCM'' labels). Each bar height is averaged over all interaction strengths and random initializations.}
    \label{fig:vqe_e_qcm}
\end{figure*}

The AIMs considered here provide a controlled setting to assess the effectiveness of this correction, since the exact ground-state energies are available from classical diagonalization. For each VQE run, we compute the absolute relative error, both for the bare variational estimate $E_{\rm VQE}$ and the infinum approximation $E_{\rm INF}$. As shown in Fig.~\ref{fig:vqe_e_qcm},
the correction estimate systematically reduces the absolute relative error introduced by both ansatz limitations and sampling noise. For the gradient-free optimal states, the correction yields a substantial reduction for both the three- and five-bath cases, indicating that higher-order moments can partially recover correlation energy that the variational state fails to capture. Gradient-based optimizations already yield smaller variational errors, particularly L-BFGS-B, yet, they also benefit from the correction. The largest relative gains appear in the larger-bath regimes. Computation of the QCM correction requires additional measurements to estimate higher-order Hamiltonian moments. For the three-bath case, this overhead is comparable to the VQE optimization cost for the L-BFGS-B optimizer, as shown in Fig.~\ref{fig:vqe_num_qcs}. In the five-bath case computing this correction is more expensive than solving the variational problem.

Since the QCM energy is not constrained to the upper bound, it can overshoot the energy for individual runs. However, the cumulant-based estimate provides, on average, a closer approximation to the exact ground-state energy. In the DMFT context, this improvement in the impurity ground-state energy is particularly valuable, as it can improve the stability of the self-consistency loop without requiring deeper ansatz for the ground state.

\subsection{Green's function construction}

To assess the accuracy with which the proposed solver can reconstruct the impurity Green's function, we estimated particle and hole Lanczos coefficients~\eqref{eq:lanczos_expansion} from Hamiltonian moments computed on the states prepared by the particle and hole excitations. These estimates were computed for every parameter set found by simulated VQE runs optimized via the L-BFGS-B algorithm, and averaged to obtain mean values for the Lanczos coefficients for a given number of bath sites and Hubbard interaction.

The resulting Lanczos coefficients were used to construct an averaged continued-fraction representation of the Green’s function for each AIM. From this representation, we then computed the corresponding density of states (DOS),
\begin{equation}
    {\rm DOS}(\omega) = -\frac{1}{\pi} {\rm Im}\{{\rm Tr}[G(\omega)]\} , 
\end{equation}
and compared it with that obtained from the exact Lanczos coefficients computed classically from the particle and hole excitations of the ground state determined via classical diagonalization.

\begin{figure*}
    \centering
    \includegraphics[width=0.47\linewidth]{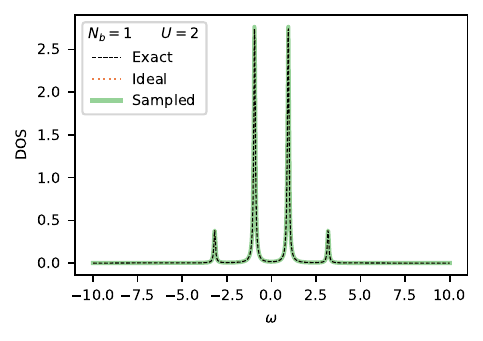}
    \includegraphics[width=0.47\linewidth]{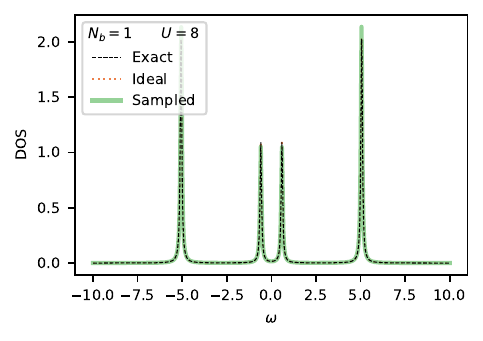}
    \includegraphics[width=0.47\linewidth]{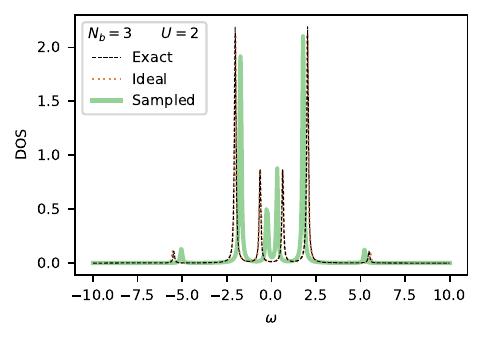}
    \includegraphics[width=0.47\linewidth]{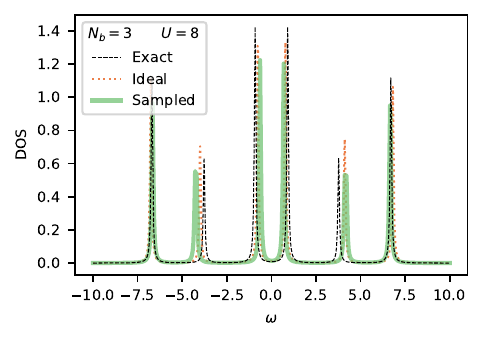}
    \includegraphics[width=0.47\linewidth]{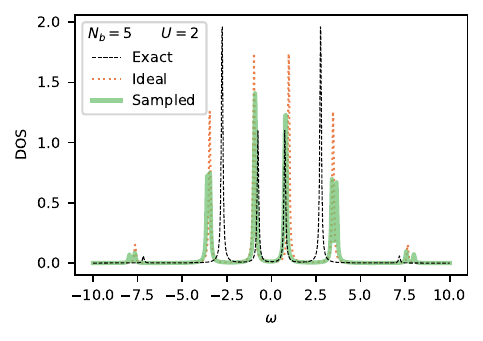}
    \includegraphics[width=0.47\linewidth]{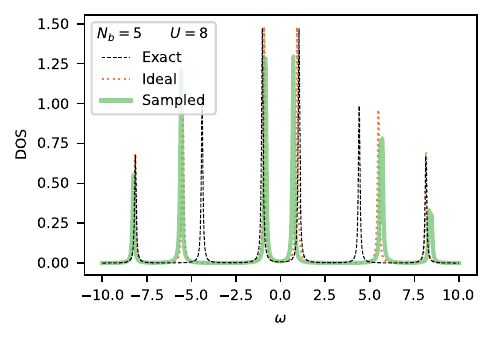}
    \caption{Comparison between DOS constructed using the proposed quantum solver (green solid lines) and that classical pipeline (black dashed lines). The shown density of states corresponds to AIMs with one (top), three (middle) and five (bottom) bath sites, with Hubbard interactions $U=2$ (left) and $U=8$ (right).}
    \label{fig:dos}
\end{figure*}

This comparison is shown in Fig.~\ref{fig:dos}. The peaks in the DOS for one-bath AIMs, shown at the top of the figure, are reproduced with very good accuracy for all values of the Hubbard interaction, as shown in the top row of the figure. As the interaction strength increases, the spectrum progressively splits into lower- and upper-Hubbard-like features, and the average DOS tracks the relative weight of these peaks across all interaction values.

Results for three-bath AIMs, shown in the middle row of Fig.~\ref{fig:dos}, capture the main features of the exact DOS over the full frequency window. This shows that, once a reasonably accurate variational ground state has been obtained, the combination of excited-state ansatz and Lanczos-based continued fractions yields a reasonable and compact representation of the impurity Green's function.

Due to the significantly higher simulation cost, the five-bath case averages were evaluated using a reduced number of configurations. Specifically, the optimization was repeated only twice per initialization to gather statistical information. These results provide valuable insight into the applicability of our ansatz and VQE-based reconstruction to larger systems. Results for this case, shown at the bottom of Fig.~\ref{fig:dos}, highlight how deviations from the true ground state lead to poor agreement with the true DOS for some values of the Hubbard interaction. Improving the construction of the Green's function for larger bath cases will prepare states closer to the true ground state for the corresponding AIMs.

The additional measurements needed to estimate the Lanczos coefficients and evaluate the continued fractions result in a quantum cost comparable to that of the VQE optimization in the three-bath case. For larger AIMs, however, this subroutine becomes the dominant cost, as the number of Pauli strings involved in the estimation of $\mu_5$ grows rapidly with the number of bath sites.

\section{Outlook}\label{sec:outlook}

In this work, we evaluated the performance of a hybrid quantum–classical solver computing the impurity Green’s function of particle–hole symmetric AIMs. The solver uses shallow symmetry-preserving ans\"atze to prepare the ground state of the model and its particle and hole excitations from the same set of variational parameters, enabling the construction of the continued-fraction representation of the Green's function via the estimation of Hamiltonian moments on the excited states.

Our comparison of different optimization strategies highlights practical cost–performance trade-offs that will become increasingly relevant as the complexity of the AIM grows. Simulations show a ``good'' variational approximation to the model's true ground state leads to a DOS, computed from the constructed Green's function, in good agreement with that obtained from a classical pipeline. As the approximation to the true ground state deteriorates with the number of bath sites due to constraints imposed on the ansatz, the fidelity of the computed DOS decreases with this number accordingly.

Therefore, improving the accuracy with which the true ground state is approximated via refinement of the ansatz and implementation of more efficient optimization strategies is paramount to further enhance the fidelity of the constructed Green’s function, particularly for larger and more complex models for which the impurity site represents highly-correlated orbitals. The proposed ansatz can be modified to prepare states with other electron numbers, enabling the treatment of systems beyond the half-filled case. Moreover, increasing the number of Givens-rotation layers will allow the ansatz to prepare more general states, albeit at the cost of increased circuit depth, which may exacerbate the barren plateau problem. An additional route to improved fidelity is adopting a more expressive ansatz in which the impurity block includes additional entangling operators (e.g., ZZ-type interactions), as explored in symmetry-preserving hardware-adaptable circuits for AIMs~\cite{jones_eb_2025}, or the bath block includes double-excitation operators. In this work, we focus on a constrained, shot-limited benchmark, and we treat the accuracy-cost tradeoff of such augmented ansätze as a natural follow-on study.

A relevant direction for future work is the integration of the quantum impurity solver into a one-shot DFT+DMFT framework, followed by its extension to a fully self-consistent scheme. Realizing this program will require improved scalability of the solver's subroutines, together with enhanced robustness via error-mitigation strategies suited to noisy intermediate-scale quantum devices. Progress along these directions would push forward towards simulations of correlated real materials with a quantum-accurate treatment of local electron interactions.

\bibliography{references}

\end{document}